# Stability of 20 Biogenic Amino Acids in Concentrated Sulfuric Acid: Implications for the Habitability of Venus' Clouds

Maxwell D. Seager[1,2,#], Sara Seager[2,3,4,5,#,*], William Bains [3,6,7] and Janusz J. Petkowski [3,8,9#]

[1] Department of Chemistry and Biochemistry, Worcester Polytechnic Institute, Worcester, MA 01609, USA

[2] Nanoplanet Consulting, Concord, MA 01742, USA

[3] Department of Earth, Atmospheric and Planetary Sciences, Massachusetts Institute of Technology, 77 Massachusetts. Avenue., Cambridge, MA 02139, USA

[4] Department of Physics, Massachusetts Institute of Technology, 77 Massachusetts. Avenue., Cambridge, MA 02139, USA

[5] Department of Aeronautics and Astronautics, Massachusetts Institute of Technology, 77 Massachusetts. Avenue., Cambridge, MA 02139, USA

[6] School of Physics & Astronomy, Cardiff University, 4 The Parade, Cardiff CF24 3AA, UK

[7] Rufus Scientific, Melbourn, Herts SG8 6ED, UK

[8] Faculty of Environmental Engineering, Wroclaw University of Science and Technology, 50-370 Wroclaw, Poland

[9] JJ Scientific, Mazowieckie, Warsaw 02-792, Poland

*Correspondence: seager@mit.edu

\# contributed equally to this work

## Abstract

Scientists have long speculated about the potential habitability of Venus, not at the 700K surface, but in the cloud layers located at 48-60 km altitudes, where temperatures match those found on Earth's surface. However, the prevailing belief has been that Venus' clouds cannot support life due to the cloud chemical composition of concentrated sulfuric acid—a highly aggressive solvent. In this work, we study 20 biogenic amino acids at the range of Venus' cloud sulfuric acid concentrations (81% and 98% w/w, the rest water) and temperatures. We find 19 of the biogenic amino acids we tested are either unreactive (13 in 98% w/w and 12 in 81% w/w) or chemically modified in the side chain only, after four weeks. Our major finding, therefore, is that the amino acid backbone remains intact in concentrated sulfuric acid. These findings significantly broaden the range of biologically relevant molecules that could be components of a biochemistry based on a concentrated sulfuric acid solvent.

## 1. Introduction

The Venus clouds may be a potential habitat for life, despite the inhospitable conditions on the surface (see e.g., Bains et al., 2023; Grinspoon and Bullock, 2007; Kotsyurbenko et al., 2021; Limaye et al., 2018; Mogul et al., 2021a; Morowitz and Sagan, 1967; Patel et al., 2021; Schulze-Makuch and Irwin, 2006; Seager et al., 2021). The cloud layers on Venus are persistent and extend vertically from around 48 to 70 kilometers above the surface (~70 km in low latitudes and ~65 km in polar latitudes), and they span the entire planet. Moreover, these cloud layers possess the essential prerequisites for supporting life (Baross et al., 2007), that is, suitable temperatures for covalent bonds, a liquid environment (comprised of cloud droplets), and an energy source in the form of sunlight (see discussion in Bains et al., 2023; and Seager et al., 2021)). The notion that life could exist in a cloud environment is supported by analogy with Earth's aerial biosphere (Amato et al., 2019; Vaïtilingom et al., 2012) and our Venus life cycle concept, which explains how microbial-type life forms could potentially remain aloft indefinitely without descending to the lower, fatally hot atmospheric layers (Seager et al., 2021).

In contrast to this promising habitability viewpoint, the Venus clouds constitute a very hostile environment for Earth life. First, the clouds are predominantly composed of liquid concentrated sulfuric acid, a highly aggressive solvent that is thought to destroy most or all biochemicals (Bains et al., 2021c). Indeed, concentrated sulfuric acid is orders of magnitude more acidic than the most acidic environments on Earth that host acid-adapted microorganisms. (For example, the hot, acidic, metal-rich Dallol pools in Ethiopia host polyextremophile organisms (Belilla et al., 2019; Gómez et al., 2019; Kotopoulou et al., 2018), but these pools are at pH ~ 1, ten orders of magnitude less acidic than concentrated sulfuric acid.) Second, there is almost no available water, and water is essential for all Earth-based life. Although the cloud droplets are composed of up to 30% water by volume (Hoffman et al., 1980; Krasnopolsky, 2015; Titov et al., 2018), the water is locked away in strong hydrogen bonds to sulfuric acid, rendering the water mostly unavailable for solvation or hydrogen bonding to other molecules. Furthermore, the atmospheric conditions outside of the droplets appear to contain minimal water vapor, resulting in an extremely arid environment (Krasnopolsky, 2015). For further details on water availability and mitigating circumstances see the works of Bains et al. (2021a, 2023); Hallsworth et al.( 2021); Mogul et al.( 2021a); and Rimmer et al.( 2021).

We advocate the notion that the droplets of highly concentrated sulfuric acid might be able to support a form of life different from that of Earth. This is based on the idea that life could use sulfuric acid instead of water as a solvent—a concept supported by survival of complex organic molecules in highly concentrated sulfuric acid (Seager et al., 2023)) and chemical reactions in highly concentrated sulfuric acid that generate a diverse set of organic molecules (Benner and Spacek, 2021; Spacek, 2021; Spacek and Benner, 2021). This perspective challenges the conventional planetary science view that only simple organic chemistry with limited functionality could be stable in concentrated sulfuric acid.

The presence of a diverse organic chemistry in concentrated sulfuric acid has long been recognized outside the realm of planetary science. In the oil refining industry, concentrated sulfuric acid is employed to process crude oil and yields a byproduct known as "red oil," which contains a wide array of organic compounds, including aromatic molecules that are stably dissolved in the concentrated sulfuric acid (Albright et al., 1972; Huang et al., 2015; Miron and Lee, 1963). Spacek and Benner (Benner and Spacek, 2021; Spacek, 2021; Spacek and Benner, 2021) demonstrated that a rich organic chemistry spontaneously arises in concentrated sulfuric acid from simple precursor molecules, such as formaldehyde. Interestingly, even gas phase CO and $CO_2$ can serve as seed molecules for this chemistry (Benner and Spacek, 2021; Spacek, 2021; Spacek and Benner, 2021), and these gases themselves could originate from photochemical processes in the atmosphere of Venus. Seager et al. (2023) conducted studies showing that nucleic acid bases, including adenine, cytosine, guanine, thymine, and uracil, as well as 2,6-diaminopurine and the "core" nucleic acid bases purine and pyrimidine, remain stable in sulfuric acid within the temperature and concentration range of the Venus clouds for at least two weeks.

Complex organic chemistry is, of course, not life, but there is no life without it. The stability of complex organic chemistry in a specific environment can be considered a prerequisite to habitability. Thus, we aim to investigate the stability of biologically relevant organic molecules in concentrated sulfuric acid.

We are motivated to study amino acids because they are one of the key fundamental building blocks of life on Earth. While there are hundreds of known amino acids and thousands of more potential ones (Brown et al., 2023), we focus on the 20 amino acids coded by Earth life's canonical genetic code, which we refer to as the 20 "biogenic" amino acids. Nine of the 20 biogenic amino acids have also been found in meteoritic material (Koga and Naraoka, 2017), which suggests a continuous supply to Venus. Our goal is to determine whether amino acids are, or are not, rapidly destroyed in concentrated sulfuric acid. We, therefore, use pure concentrated sulfuric acid for this foundational test, leaving the trace components in the Venus clouds and related chemistry for future work.

## 2. Materials and Methods

We purchased a set of L-amino acids from Millipore-Sigma, in the form of a set of biogenic L-amino acids with ≥ 98% purity (catalog number LAA21-1KT). Cysteine was in the form of cysteine hydrochloride and lysine in the form of lysine monohydrochloride. L-Methionine sulfoxide (catalog number M1126), L-Methionine sulfone (catalog number M0876), and *S*-sulfo-L-cysteine (catalog number C2196) all with purity ≥ 98% were purchased from Millipore-Sigma. L-Cysteine sulfinic acid (purity ≥ 98%; catalog number sc-203620), L-Cysteic acid (purity ≥ 98%; catalog number sc-485621), and *O*-Sulfo-L-serine (purity > 99%; catalog number sc-295959), were purchased from Santa Cruz Biotechnology, Inc. Iron (II) oxide (FeO) (purity ≥ 99.6%; catalog number 400866) and formic acid (purity ≥ 98%; catalog number 33015) were purchased from Sigma Aldrich. The compounds were used without further purification. We used $D_2SO_4$ from ACROS Organics (sulfuric acid-$d_2$ for NMR, 98 wt.% in $D_2O$, 99.5+ atom % D) and $D_2O$ (deuteration degree min 99.9%) from MagniSolv.

We prepared our NMR samples by dissolving 25 to 80 mg of the amino acids into 500 mL of solvent $D_2SO_4$ in $D_2O$ in glass vials. We used 10 to 40 mg of compounds for the 1D $^1H$ and $^{13}C$ NMR. We used 50 to 80 mg for some of the 2D NMR. When required, we heated sealed glass vials in a hot water bath (~80 ºC for a few minutes) to promote dissolution of the compounds. We transferred the solution to 5 mm NMR tubes and stored the tubes for 12 to 18 hours before NMR measurements. After NMR measurements, we stored the solutions in the NMR tubes, where the storage room temperature varied from about 18 to 24 ºC.

To acquire NMR data, we used a Bruker Avance III-HD 400 MHz spectrometer equipped with a Prodigy liquid nitrogen cryoprobe (BBO) at 25 °C. We acquired 1D $^1H$, $^{13}C$, and 2D $^1H$-$^{13}C$ HMQC NMR spectra to confirm the structures and, hence, stability of the compounds in 98% w/w and 81% w/w $D_2SO_4$ in $D_2O$. In all cases we locked on $D_2SO_4$. The $D_2SO_4$ peak is at 11.46 +/- 0.02 ppm in 98% w/w $D_2SO_4$ and at 11.99 +/- 0.02 ppm in 81% w/w $D_2SO_4$.

We used MNova software (Mestrelab Research) to process and analyze the NMR data (Willcott, 2009). The original data for all NMR experiments are available for download as Supplementary Datasets from Zenodo at https://zenodo.org/record/8381013.

## 3. Results: Stability and Reactivity of Biogenic Amino Acids in Concentrated Sulfuric Acid

Our major finding is that the backbone of the amino acid molecule remains intact in concentrated sulfuric acid. We studied the 20 biogenic amino acids in concentrated sulfuric acid at the concentration range found in Venus' clouds (81% w/w and 98% w/w, the rest water) at room temperature. We found that 19 of the biogenic amino acids are either unreactive or chemically modified in the side chain only after four weeks in 98% w/w or 81% w/w concentrated sulfuric acid at room temperature and pressure. We note that for all of the biogenic amino acids, the amino, and carboxyl groups are fully protonated, as they are in water with acidic pH (see SI).

We found that eleven of the biogenic amino acids are stable in both 81% w/w and 98% w/w sulfuric acid. Nine are reactive, eight of which are stable to further change after rapid formation of side-chain derivatives in concentrated sulfuric acid. We now describe details in the text, Table 1, and Figures 1-4.

Out of the nine amino acids that are reactive in concentrated sulfuric acid, two are chemically modified in a similar way in concentrated sulfuric acid as they are in water (asparagine and glutamine). In 81% w/w sulfuric acid, both asparagine and glutamine are slowly converted to aspartic acid (in one week) and glutamic acid (in four weeks), a hydrolytic reaction that also happens more slowly in water (Figure S27). In 98% w/w concentrated sulfuric acid, asparagine and glutamine are unchanged (*i.e.*, not converted into aspartic acid and glutamic acid), presumably due to the much lower water activity.

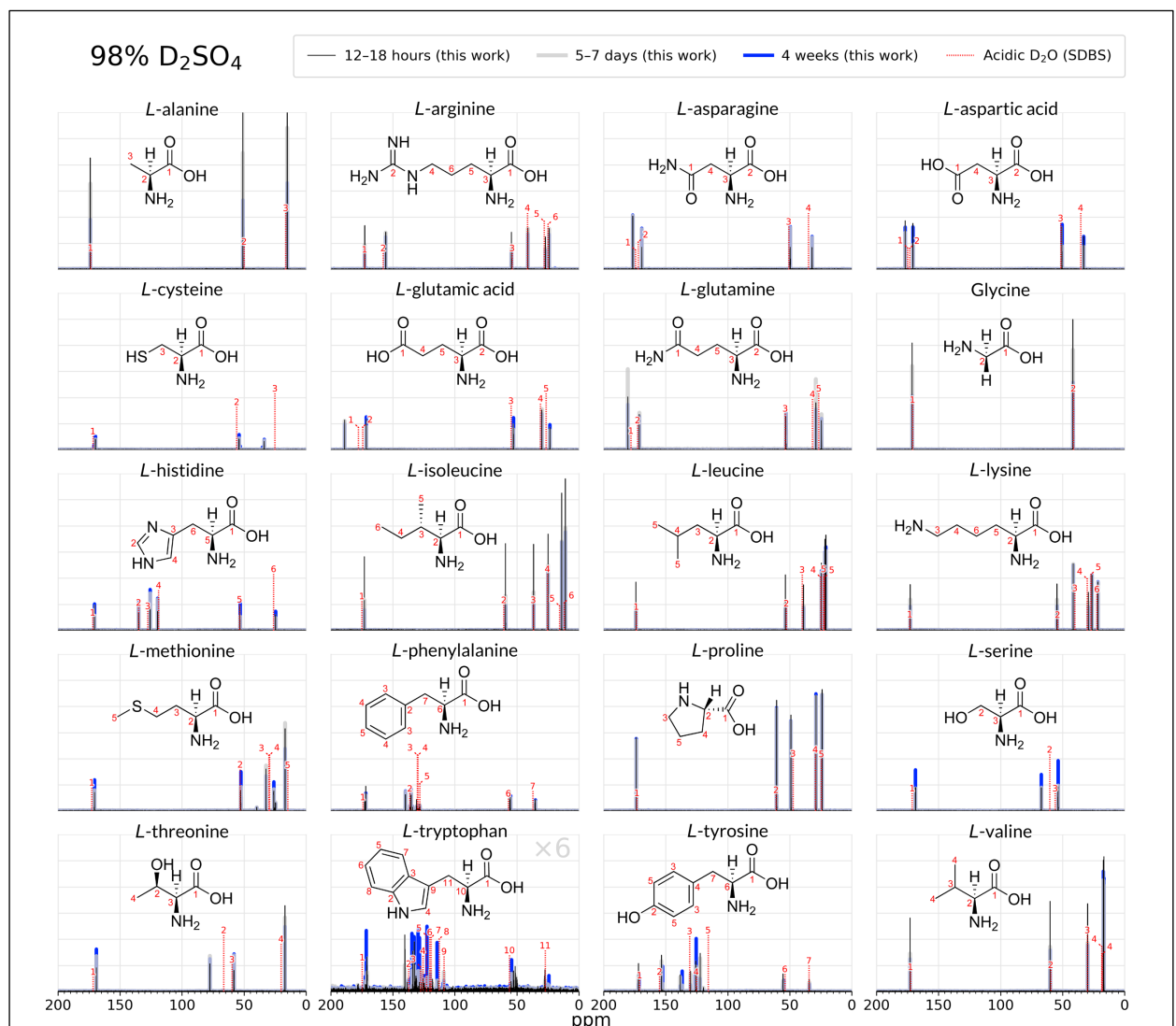

**Figure 1.** Comparison of the 1D $^{13}$C NMR spectra of 20 biogenic amino acids after 12-18h (black), 5-8 days (grey), and after four weeks (blue) of incubation in 98% $D_2SO_4$/2% $D_2O$ (by weight). Thirteen amino acids (arginine, histidine, lysine, aspartic acid, glutamic acid, asparagine, glutamine, glycine, proline, alanine, isoleucine, leucine, valine) remain stable and unchanged for four weeks in 98% w/w $D_2SO_4$, as shown by the virtually identical peaks in the four week spectra and the 12-18h spectra. A comparison with NMR peaks for each compound in acidic $D_2O$ (dashed lines; Li et al., 2020; Saito et al., 2006) further confirms the structural integrity of the amino acids and additionally shows that the overall molecular structure is not affected by the concentrated sulfuric acid solvent. The side chains of amino acids methionine, cysteine, serine, threonine, phenylalanine, tyrosine and tryptophan undergo rapid chemical modification, as shown by emergence of additional NMR spectral peaks. For this set, with the exception of tryptophan, the amino acid backbone is intact, as illustrated by the unchanged NMR spectral peaks belonging to the α-carbon and the carboxylic group. Note that tryptophan intensities in sulfuric acid are multiplied by 6 so the peaks are visible at the scale of the figure. SDBS is the Spectral Database for Organic Compounds (Saito et al., 2006). See the SI for more detailed NMR figures for each molecule.

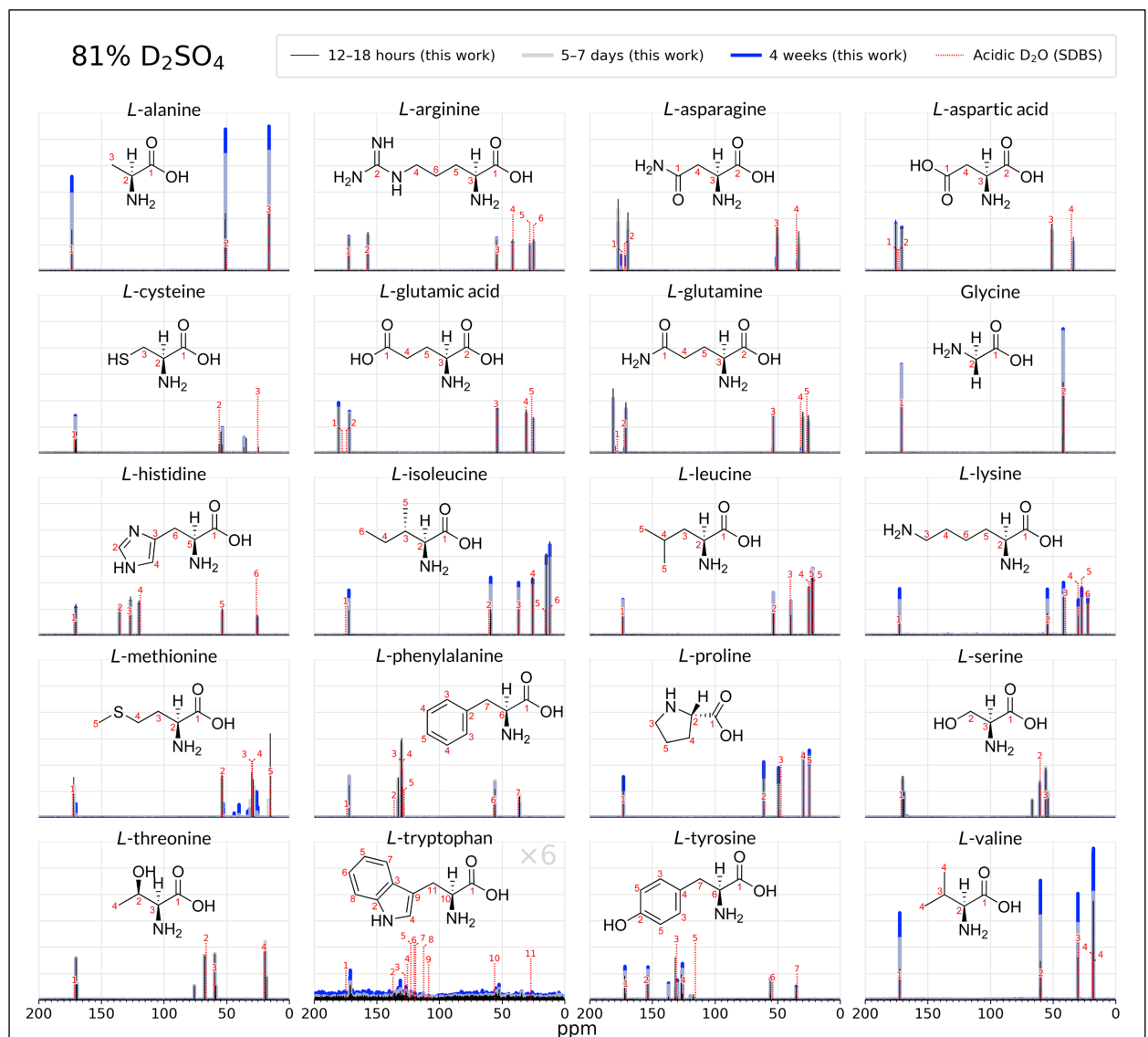

**Figure 2.** Comparison of the 1D $^{13}$C NMR spectra of 20 biogenic amino acids after 12-18h (black), 5-8 days (grey) and after four weeks (blue) of incubation in 81% D$_2$SO$_4$/2% D$_2$O (by weight). Twelve amino acids (arginine, histidine, lysine, aspartic acid, glutamic acid, glycine, proline, alanine, isoleucine, leucine, valine, phenylalanine) remain stable and unchanged for four weeks in 81% w/w D$_2$SO$_4$, as shown by the virtually identical peaks in the four week spectra and the 12-18h spectra. A comparison with NMR peaks for each compound in acidic D$_2$O (dashed lines; Li et al., 2020; Saito et al., 2006) further confirms the structural integrity of the amino acids and additionally shows that the overall molecular structure is not affected by the concentrated sulfuric acid solvent. The amino acids methionine, cysteine, serine, threonine, asparagine, glutamine, tyrosine and tryptophan side chains undergo rapid chemical modification, as shown by emergence of additional NMR spectral peaks. For this set, with the exception of tryptophan, the amino acid backbone is intact, as illustrated by the unchanged NMR spectral peaks belonging to the α-carbon and carboxylic group. Note that tryptophan intensities in sulfuric acid are multiplied by 6 so the peaks are visible at the scale of the figure. SDBS is the Spectral Database for Organic Compounds (Saito et al., 2006). See the SI for more detailed NMR figures for each molecule.

group to an oxygen or sulfur atom) in concentrated sulfuric acid, and two are sulfonated (addition of an $SO_3H$ group to a carbon atom), both of which are chemical modifications that cannot happen in pure water.

Serine and threonine are sulfated at the hydroxyl alcohol functional group (Figure S32), which is consistent with previously reported NMR spectra of sulfoserine and sulfothreonine (Rose et al., 1994). The sulfation of serine and threonine is less efficient in 81% w/w than in 98% w/w sulfuric acid. This conclusion is supported by the two populations of $^{13}C$ carbon peaks in the NMR spectra in 81% w/w concentration that correspond to the two populations of serine and threonine – sulfated and native. This is an expected result, as sulfation of a hydroxyl is a dehydration reaction, and 98% w/w concentrated sulfuric acid is a more powerful dehydrating agent than 81% w/w concentrated sulfuric acid. We note that serine and threonine are deliberately sulfated by some life on Earth as a posttranslational modification of proteins to modify biological functions of target proteins (Medzihradszky et al., 2004).

In cysteine, the thiol functional group is sulfated in concentrated sulfuric acid and yields *S*-sulfocysteine (Figure S28). This conclusion is supported by the comparison of four-week incubated cysteine to the $^1H$ and $^{13}C$ NMR spectra of native *S*-sulfocysteine in concentrated sulfuric acid (Figure S28). Methionine is likely demethylated to homocysteine and subsequently sulfated to yield *S*-sulfohomocysteine in concentrated sulfuric acid (Figure S31). Such demethylation reaction of methionine has previously been proposed to occur in concentrated sulfuric acid (Andrews and Bruce, 1951).

We now turn to the final three amino acids that are reactive in concentrated sulfuric acid. Phenylalanine appears to be sulfonated at multiple locations in the ring in 98% w/w sulfuric acid within one day, as shown by the additional carbon peaks in the "aromatic region" of the $^{13}C$ NMR spectra. In contrast, in 81% w/w sulfuric acid, the phenylalanine structure remains unchanged for at least four weeks. The modification of the phenylalanine aromatic ring in 98% w/w is likely not an oxidation, as oxidation of aromatics is not known to happen in concentrated sulfuric acid (Liler, 1971).

Tyrosine appears to undergo several chemical changes over time in 98% w/w concentrated sulfuric acid, likely due to sulfation of the hydroxyl alcohol functional group (Liler, 1971) and/or sulfonation of the phenyl ring, producing multiple populations of modified molecules, as illustrated by the appearance of a multitude of additional peaks on the $^{13}C$ NMR spectrum. The early studies on the chemical reactivity of proteins and amino acids in 98% w/w sulfuric acid also support the sulfation and sulfonation of the phenyl ring of tyrosine (Habeeb, 1961; Reitz et al., 1946). Tyrosine in 81% w/w concentrated sulfuric acid likely becomes sulfated at the hydroxyl alcohol functional group and possibly sulfonated as well. We note that tyrosine in water is oxidized (e.g., (Burzio and Waite, 2002; Recky et al., 2021); whether such oxidation in concentrated sulfuric acid happens remains to be ascertained.

Only one of the chemically modified amino acids is highly reactive in concentrated sulfuric acid. Tryptophan undergoes rapid and complex reactivity that likely involves the formation of highly diverse cross-linked species[1] as, for example, noted in passing by Ramachandran and McConnell (1955). We note that in water tryptophan can readily be oxidized by reactive oxygen species such as singlet oxygen that can be generated from atmospheric oxygen by UV light or reaction with metals such as iron, to react into a variety of species (Bellmaine et al., 2020; Simat and Steinhart, 1998).

The remaining 11 biogenic amino acids (alanine, arginine, aspartic acid, glutamic acid, glycine, histidine, isoleucine, leucine, lysine, proline, and valine) show no detectable reactivity and degradation of the tested compound after incubation in concentrated sulfuric acid for at least four weeks in both 98% w/w and 81% w/w concentrated sulfuric acid at room temperature. These 11 biogenic amino acids are also stable in water. We demonstrate their stability by comparing their $^{13}$C NMR spectral peak shifts in concentrated sulfuric acid with published literature values obtained in acidic water (see the SI, Tables S1-S20) (Saito et al., 2006) and show they are the same and do not change with time (Figure 1, Figure 2). We further confirm the amino acid structures via $^{1}$H NMR and $^{1}$H-$^{13}$C 2D HMQC NMR (see SI, Figures S1-S4).

## 4. Discussion

Our major finding is that the backbone structure of 19 of the 20 biogenic amino acids is unperturbed in concentrated sulfuric acid (the exception being tryptophan). Eleven are stable in concentrated acid; the other eight are modified only in the side chain.

Concentrated sulfuric acid's chemical properties differ significantly from that of water, so much so that our results may appear unexpected. Our findings help to challenge the prevailing misconception in the astrobiology and biology communities that organic chemicals are uniformly unstable in concentrated sulfuric acid[2]. In fact, experiments demonstrating stability of organic compounds in concentrated sulfuric acid research date back to the 1920s, though unconnected to Venus clouds. Examples of such early research include stability and reactivity studies of proteins (Bischoff and Sahyun, 1929; Habeeb, 1961; Reitz et al., 1946), lipids (Steigman and Shane, 1965), and various other organics, for example, Albright et al. (1972) and Miron and Lee (1963). See also the work of Bains et al. (2021b, 2021c) for a recent review of reactivity and stability of organic chemicals in concentrated sulfuric acid.

The Venus clouds are expected to have trace components and be different from pure "test tube" conditions. Extensive in situ and Earth-ground based observations over the past several decades show a variety of gases, metals, and other compounds detected in the cloud layers and cloud particles (see the recent review article of Petkowski et al., 2023) and the original work of Andreichikov (1987b and 1987a); Mogul et al. (2021b); Petrianov et al. (1981) and; Zolotov et al. (2023)). Before addressing reactivity in complex mixtures, we need preliminary fundamental experiments, hence this work's focus on pure concentrated sulfuric acid. Some of the trace species in the Venusian clouds, however, may react with amino acids. If there is reactivity, we want to acknowledge that, for life to function, there is a crucial balance between biochemical's reactivity and stability. Life relies on chemicals that are mostly on the verge of spontaneous chemical reactivity, so that minimal catalysis can allow the reactions of metabolism. Extremely stable compounds, such as fluorocarbon "forever chemicals", tend to be toxic because biochemistry cannot process them (Brunn et al., 2023). Testing with trace species is a topic to be expanded upon in future work. To initiate such studies, we tested the stability of two key amino acids, alanine and glycine, in the presence of two key contaminants, CO (released from formic acid) and FeO, at room temperature, and found that the amino acids are stable under these conditions as well as in pure concentrated sulfuric acid (tested for up to one week) (see SI Section S5 and Figures S33-S35 for details).

---

[1] We note that the $O_2$ from the atmosphere could contribute to the destruction of tryptophan in concentrated sulfuric acid.
[2] Perhaps such misconceptions stem from the improper extrapolation of the results of a popular experiment of sugar's reactivity with concentrated sulfuric acid (Pines et al., 2012).

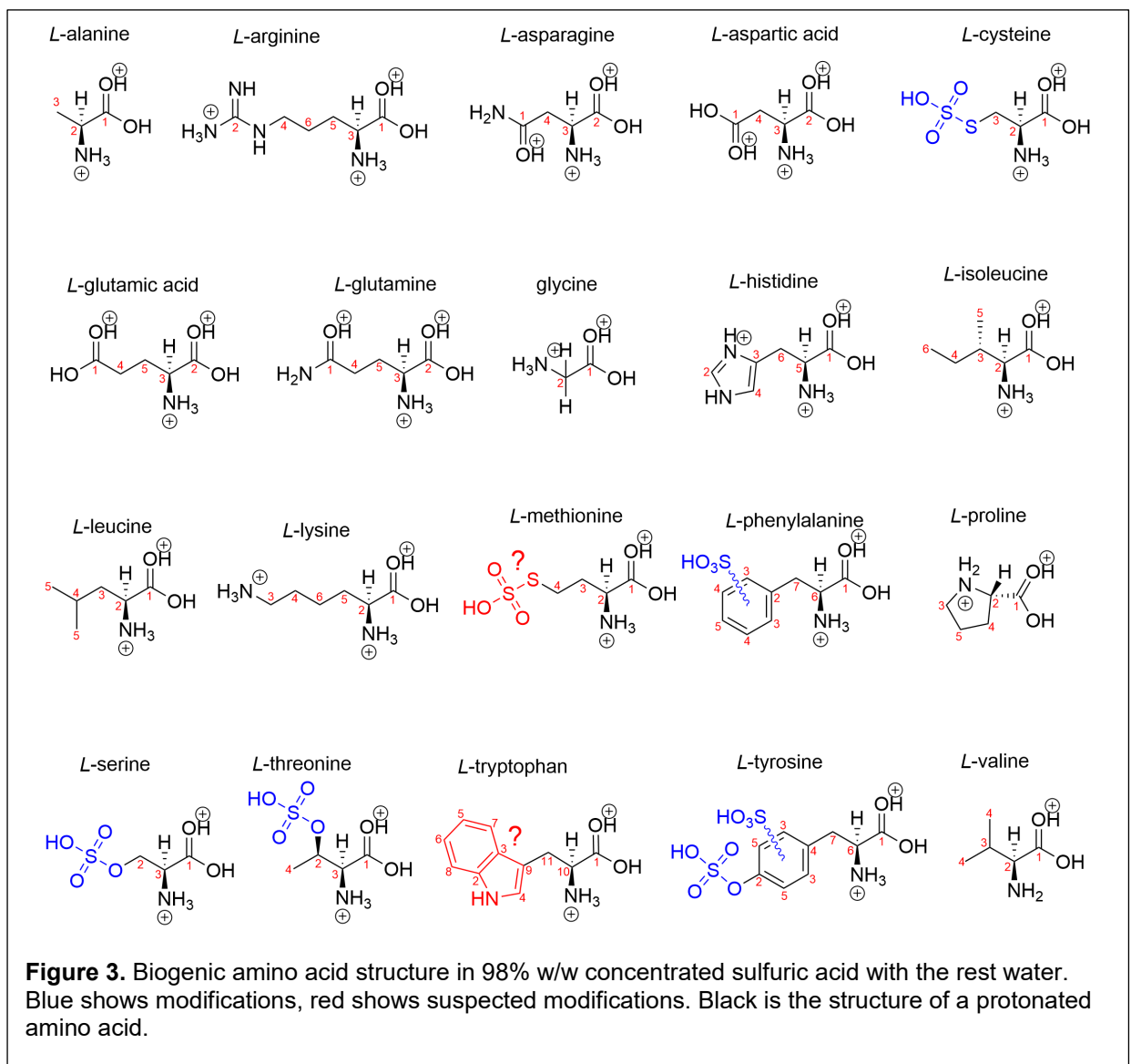

**Figure 3.** Biogenic amino acid structure in 98% w/w concentrated sulfuric acid with the rest water. Blue shows modifications, red shows suspected modifications. Black is the structure of a protonated amino acid.

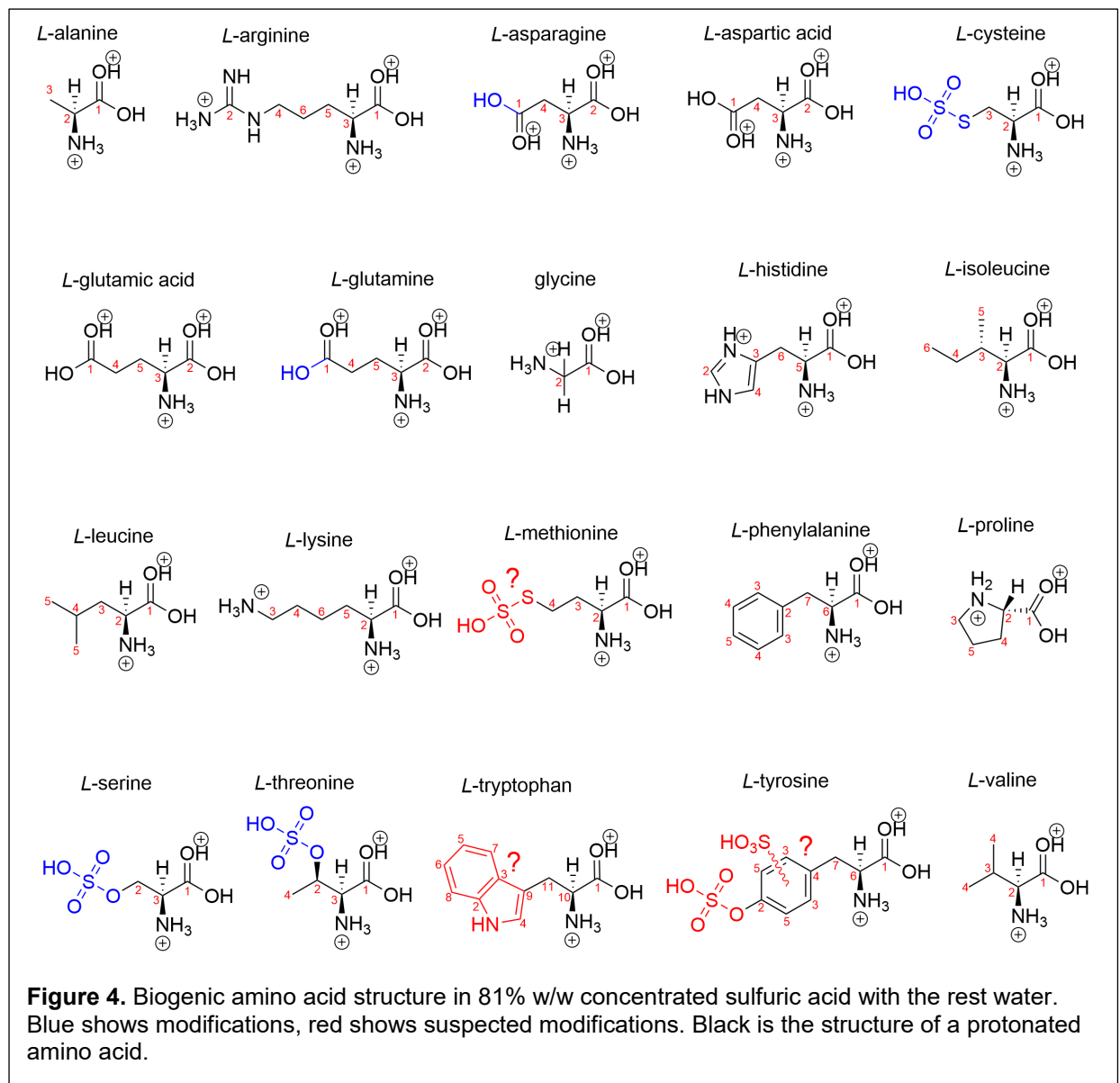

**Figure 4.** Biogenic amino acid structure in 81% w/w concentrated sulfuric acid with the rest water. Blue shows modifications, red shows suspected modifications. Black is the structure of a protonated amino acid.

Regarding the modification of the amino acid side chains seen in our experiments, our own life routinely modifies side chains of amino acids. Out of the 20 amino acids in our protein-building repertoire only the side chains of small hydrophobic amino acids alanine, valine, leucine, and isoleucine are not modified by life. The rest are often modified (a phenomenon called post-translational modification of amino acids). Furthermore, often one amino acid can be modified in many different ways (e.g., lysine can be either methylated or acetylated or even phosphorylated). Such amino acid modifications give life an ability to modify protein function, including the ability to give existing proteins entirely new functions in the cell.

There are a few key points in relating our amino acid studies in concentrated sulfuric acid to Venus cloud habitability. There is the question of which amino acids any hypothetical Venus life might use. We chose to test the stability and reactivity of the 20 biogenic amino acids used by life on Earth as they are a well-defined set of compounds that our life uses. Nine of the biogenic amino acids have also been identified in meteoritic material (Koga and Naraoka, 2017); seven of these fall into the category of amino acids unreactive and two are are sulfated (serine and threonine) and then are stable to further reactivity in concentrated sulfuric acid. Therefore, presumably a small but steady supply of some amino acids could be delivered via meteoritic infall to the Venus clouds. There are 500 known amino acids and thousands of more possibilities (Brown et al., 2023). Our work shows the backbone of the amino acid is stable in concentrated sulfuric acid, pointing towards a large variety of amino acids to be stable in concentrated sulfuric acid.

Our work also informs the possible origins of life on Venus, if life exists there. In one scenario, life may have originated in water oceans hypothesized to have existed on Venus' surface for up to billions of years, before evaporating (Way et al., 2016; Way and Del Genio, 2020). As water became very scarce, life would have adapted its biochemistry from water to a concentrated sulfuric acid environment. In an opposing scenario, Venus may always have been too warm to host water oceans (Turbet et al., 2021) and may have had concentrated sulfuric acid in its clouds as a dominant liquid for most of its geological history, in which case prebiotic chemistry would have to generate life in concentrated sulfuric acid. This second scenario might be thought implausible, if the traditional view of concentrated sulfuric acid as inimical to complex chemistry is held. Our work shows that either scenario could be plausible routes to the origin of life, as we show that eleven of our twenty amino acids are unmodified in both 98% w/w and 81% w/w sulfuric acid, so that a set of unreactive amino acids to utilize is readily available in concentrated sulfuric acid no matter how life ends up there.

Our goal was to check whether amino acids are or are not rapidly destroyed in concentrated sulfuric acid. Modifications that do occur to amino acid side chains in concentrated sulfuric acid generally happen quickly, within hours to days. We stopped monitoring the reactivity of amino acids in concentrated sulfuric acid after four weeks of incubation at room temperature, as our experiments showed no further change in stability. (The one exception was tryptophan, which degraded completely in concentrated sulfuric acid; and after four weeks, new dominant chemical species emerged from the reaction's unknown products). We note that not all terrestrial biochemicals are stable in water. For example, asparagine spontaneously converts to aspartic acid in proteins on a timescale of months even at neutral pH (Yang and Zubarev, 2010). Other amino acids, depending on the reaction conditions, may oxidize on a timescale of hours to days (cysteine, methionine, tryptophan, and tyrosine) (Grassi and Cabrele, 2019). Longer term stability studies of amino acids in concentrated sulfuric acid may be of interest for future work.

A final key point regarding the connection between our results and Venus' habitability has to do with further questions that include racemization of amino acid chiral centers

(e.g., Bada, 1985), the possibility of utilization of amino acids (or similar building blocks) in large polymeric structures, and, as mentioned, amino acid stability in non-pure sulfuric acid droplets as the Venus cloud particles likely are. These are also questions for future work.

Concentrated sulfuric acid's properties differ vastly from aqueous or diluted acid solutions, which challenges typical organic and biochemistry assumptions. Studying organic chemistry in pure sulfuric acid requires a direct, unbiased approach, without prior preconceptions and assumptions, as highlighted by Spacek and Benner's works (Benner and Spacek, 2021; Spacek, 2021; Spacek and Benner, 2021). Our work furthers this concept by presenting unexpected findings regarding the reactivity and stability of amino acids—one of potential life's building blocks in this unique solvent.

Venus, our neighboring planet, lies conveniently close, which allows us to directly probe its cloud particles through space missions. In the 1970s and 1980s NASA's Pioneer Venus in situ missions and several Soviet descent craft and landers studied the clouds but did not unambiguously determine the composition of all types of cloud particles or search for the presence of organic chemicals. Presently, NASA and ESA have plans to dispatch missions to Venus by the end of this decade (Garvin et al., 2022; de Oliveira et al., 2018; Smrekar et al., 2022), and Rocket Lab plans to launch a small mission in 2025 (French et al., 2022). Ultimately, a sample return from the Venusian atmosphere may be necessary to robustly ascertain the presence of life, if indeed life exists there (Limaye and Garvin, 2023; Schulze-Makuch and Irwin, 2002; Seager et al., 2022; Shibata et al., 2017).

We are at the dawn of a new branch of Astrobiology and a new branch of organic chemistry. We close with a call to action to study organic chemistry in alternative solvents from water, which is crucial for the true understanding of the extent of the habitability of the Galaxy.

## Acknowledgements

We thank the MIT Department of Chemistry Instrumentation Facility Director Walter Massefski and NMR Consultant Bruce Adams. We thank Lauren Herrington for preparation of Figures 1, 2, S5, and S6. We thank John Grimes and Sagi Ravid for useful discussions. We thank the reviewers for helpful comments which improved the paper.

## Data Availability Statement

Original data are deposited in Zenodo data repository at https://zenodo.org/record/8381013.

## Author Contributions

M.D.S., S.S., J.J.P. designed research; S.S., M.D.S. J.J.P. performed research; J.J.P., M.D.S., S.S. analyzed data; S.S., M.D.S. J.J.P., W.B. edited the paper; S.S., M.D.S. J.J.P. wrote the paper.

## Competing Interests

The authors declare no competing interests.

## Funding

This work was partially funded by MIT and Nanoplanet Consulting LLC.